\documentclass[pra,twocolumn,superscriptaddress]{revtex4}
\usepackage{amsmath}
\usepackage{amssymb}
\usepackage{graphicx}

\newcommand{\Hop}{\hat{H}} 
\newcommand{\Dop}{\mathcal{D}}
\newcommand{\alop}{\hat{\alpha}}
\newcommand{\aldop}{\hat{\alpha}^{\dagger}}
\newcommand{\aop}{\hat{a}}
\newcommand{\adop}{\hat{a}^{\dagger}}
\newcommand{\bop}{\hat{b}}
\newcommand{\bdop}{\hat{b}^{\dagger}}
\newcommand{\cop}{\hat{c}}
\newcommand{\cdop}{\hat{c}^{\prime}}
\newcommand{\hb}{\mathbf{h}}
\newcommand{\gb}{\mathbf{g}}
\newcommand{\rhoop}{\hat{\rho}}

\newcommand{\Gm}{\mathbf{G}}
\newcommand{\LMP}{\mathbf{\Lambda}^+}
\newcommand{\LMM}{\mathbf{\Lambda}^-}
\newcommand{\lmp}{{\pmb{\lambda}_P}} 
\newcommand{\lmpc}{{\pmb{\lambda}_P^{\ast}}} 
\newcommand{\lmt}{{\tilde{\pmb{\lambda}}}}
\newcommand{\lmtc}{{\tilde{\pmb{\lambda}}^{\ast}}}

\newcommand{\Omegam}{\mathbf{\Omega}}

\newcommand{\sgx}{\hat{\sigma}^x}
\newcommand{\sgy}{\hat{\sigma}^y}
\newcommand{\sgz}{\hat{\sigma}^z}
\newcommand{\sgp}{\hat{\sigma}^+}
\newcommand{\sgm}{\hat{\sigma}^-}

\newcommand{\Am}{\mathbf{A}}
\newcommand{\Bm}{\mathbf{B}}
\newcommand{\Cm}{\mathbf{C}}
\newcommand{\Dm}{\mathbf{D}}
\newcommand{\Em}{\mathbf{E}}
\newcommand{\Fm}{\mathbf{F}}
\newcommand{\Tm}{\mathbf{T}}
\newcommand{\Rm}{\mathbf{R}}
\newcommand{\Pm}{\mathbf{P}}
\newcommand{\Mm}{\mathbf{M}}
\newcommand{\Jm}{\mathbf{J}}
\newcommand{\Xm}{\mathbf{X}}
\newcommand{\Ym}{\mathbf{Y}}
\newcommand{\Zm}{\mathbf{Z}}
\newcommand{\Wm}{\mathbf{W}}
\newcommand{\Om}{\mathbf{O}}

\newcommand{\Qm}{\mathbf{Q}}

\newcommand{\Kp}{\mathbf{K}^+}
\newcommand{\Km}{\mathbf{K}^-}
\newcommand{\Kpm}{\mathbf{K}^{\pm}}

\newcommand{\Pop}{\mathcal{P}}
\newcommand{\Lop}{\mathcal{L}}
\newcommand{\PA}{\mathcal{A}}
\newcommand{\PB}{\mathcal{B}}
\newcommand{\PC}{\mathcal{C}}

\newcommand{\hc}{{\rm H.c.}} 
\newcommand{\im}{{\rm i}} 
\newcommand{\tr}{{\rm tr}}

\newcommand{\id}{{\textbf{1}}}

\newcommand{\sutd}{EPD Pillar, Singapore University of Technology and Design, 8 Somapah Road, 487372 Singapore} 
\newcommand{\zz}{Zhengzhou Information Science and Technology Institute, Zhengzhou 450004, China}

\begin{document}

\title{Analytical solutions for a boundary driven $XY$ chain}
\author{Chu Guo}
\affiliation{\zz}
\author{Dario Poletti} 
\affiliation{\sutd} 

\begin{abstract}    
We study non-interacting fermionic systems dissipatively driven at their boundaries, focusing in particular on the case of a non-number-conserving Hamiltonian, which for example describes an $XY$ spin chain. We show that despite the lack of number conservation, it is possible to convert the problem of calculating the normal modes of the master equations and their corresponding rapidities, into diagonalizing simply an $L\times L$ tridiagonal bordered $2-$Toeplitz matrix, where $L$ is the size of the system. Such structure of matrix allows us to further reduce the problem into solving a scalar trigonometric non-linear equation for which we also show, in the case of an Ising chain, exact analytical explicit, and system size independent, solutions. 

\end{abstract}

\date{\today}
\maketitle

\address{} 

\vspace{8mm}

\section{introduction}
Quantum systems in contact with an environment are a very important topic which concerns many branches of physics, including quantum optics\cite{Gardiner2000}, quantum thermodynamics \cite{BenentiWhitney2017}, quantum computing \cite{NielsenChuang2000} and more. A particularly important class of open quantum systems is that of boundary driven systems, where the system is coupled to the environment only at the extremities. The boundary dissipative coupling drives the system towards a non-equilibrium steady state (NESS) which usually present a non-vanishing current which depends significantly on the properties of the system. For this reason, boundary dissipatively driven quantum systems are particularly important to study quantum transport.

Similarly to the case of closed quantum systems, for open quantum system it is useful to have analytical solutions to guide the physical understanding of more complex systems. However, for open quantum systems the number of known analytical solutions is limited. In the following we will consider only open quantum systems whose dynamics is described by a master equation in Lindblad form \cite{GoriniSudarshan1976, Lindblad1976}, and for which the generator of the evolutoin is called the Lindbladian. An exact matrix product ansatz can be constructed for the boundary driven $XXZ$ chain in some regimes of boundary dissipative driving \cite{Prosen2011a, Prosen2011b, Prosen2014, Popkov2013b, Popkov2014, Prosen2015}. Furthermore, for a boundary driven $XX$ chain also in the presence of dephasing, a cleverly designed ansatz has been used to exactly calculate the one-point and two-point correlation functions \cite{Znidaric2010}. The spectrum of the Lindblad operator (Lindbladian) of the tight-binding fermionic chain (or $XX$ chain) in a dephasing environment has also been exactly computed by mapping it into a Hubbard chain with imaginary interaction strength \cite{MedvedyevaProsen2016}. 

A different class of analytically solvable open quantum many-body systems belongs to the class of quadratic bosonic or fermionic systems. For a boundary driven non-interacting bosonic, fermionic and $XX$ chain, correlation functions have been analytically computed \cite{Manzano2012, Manzano2013}. In a seminal work in $2008$ \cite{Prosen2008}, Prosen showed that diagonalizing the Lindbladian of any quadratic fermionic system can be reduced to diagonalizing a $4L\times4L$ antisymmetric matrix, which can be further reduced to diagonalizing a $2L\times 2L$ generic matrix \cite{Prosen2010}, that is a matrix with no obvious symmetries and for which it is difficult to find explicit analytical exact expressions. He also provided a perturbative expression for the relaxation gap, that is the real part of the slowest decaying modes, for a boundary driven Ising chain, obtained in the limit of large $L$. Similar calculations were developed for non-interacting bosons \cite{ProsenSeligman2012}.      

In our previous work \cite{GuoDario2017}, we showed that for number conserving quadratic systems (be them bosonic, fermionic or made of spins), finding that the rapidities (eigenvalues of the Lindbladian) and the decay modes could be reduced to the problem of diagonalizing an $L\times L$ matrix which could be of special form, i.e. a bordered Toeplitz matrix, which has known analytical solutions \cite{Yueh2005, Kouachi2006, Willms2008}. This allowed us to find explicit analytical solutions.     

However, for the case of non-number conserving Hamiltonians, the method described in \cite{GuoDario2017}, would result in diagonalizing an $2L\times 2L$ block bordered Toeplitz matrix, for which we could not find analytical solutions. Building on this approach, we now address the problem of solving a boundary driven $XY$ chain. We found that the problem contains another symmetry which, when exploited, would allow to turn the problem into diagonalizing an $L\times L$ tridiagonal bordered $2-$Toeplitz matrix, which can be reduced to solving a scalar trigonometric non-linear equation. Moreover we give explicit solution for the special case of an Ising chain with transverse baths, and we show that the relaxation gap is independent of the system size $L$.

We here summarize the main steps discussed in the paper, highlighting the key equations. In Sec.\ref{sec:model} we introduce the dissipatively boundary driven model we study. In Sec.\ref{sec:solvemaster} we show how to reduce the problem of diagonalizing the Lindbladian to diagonalizing a $2L\times 2L$ matrix. Then in Sec.\ref{sec:observable} we show that, in ordering to compute observables, it is sufficient to solve a Lyapunov equation (Eq.(\ref{eq:pxxp})) which can efficiently be solved numerically. In Sec.\ref{sec:solveXY}, we apply our approach to the case of a boundary driven $XY$ chain, for which we use the symmetries in Eq.(\ref{eq:eigenPsub}), to turn the problem into solving an $L\times L$ bordered $2-$Toeplitz matrix ($\Qm^{\pm}$ in Eq.(\ref{eq:Qpm})), which can be turned into solving the trigonometric equation (\ref{eq:lambda}). For the special case of an Ising chain, we find $L-$independent analytical solutions given in Eq.(\ref{eq:solutions}). In Sec.\ref{sec:summary} we draw our conclusions.

\section{model}\label{sec:model}
We consider an open quantum system of L sites with fermionic particles. Its dynamics is described by the quantum Lindblad master equation \cite{GoriniSudarshan1976, Lindblad1976} with Lindbladian $\Lop$
\begin{align}\label{eq:Lindblad}
\frac{d\rhoop}{dt} = \Lop(\rhoop) = -\im \left[\Hop, \rhoop\right] + \Dop(\rhoop).
\end{align}
Here $\rhoop$ is the density operator of the system, $\Hop$ is the Hamiltonian, and the dissipator $\Dop$ describes the dissipative part of the evolution.
We consider the Hamiltonian
\begin{align}\label{eq:Ham}
\Hop = \sum_{i,j=1}^L\hb_{i,j}\aldop_i\alop_j + \frac{1}{2}\sum_{i,j=1}^L\gb_{i,j}\aldop_i\aldop_j + \frac{1}{2} \sum_{i,j=1}^L\gb_{j,i}^{\ast}\alop_i\alop_j, 
\end{align}
where $\aldop_j (\alop_j)$ creates (annihilates) one fermion on site $j$. $\hb$ is an $L\times L$ Hermitian matrix, and $\gb$ is an $L\times L$ anti-symmetric matrix satisfying $\gb^t = -\gb$. The dissipative part is given by
\begin{align}\label{eq:dissipator}
\Dop(\rhoop) = &\sum_{i,j=1}^L \left[\LMP_{i,j}\left(\aldop_i\rhoop\alop_j-\alop_j\aldop_i\rhoop\right)
\right. \nonumber \\ &+ \left. \LMM_{i,j}\left(\alop_i\rhoop\aldop_j-\aldop_j\alop_i\rhoop\right) + \hc  \right],
\end{align}
where $\LMP$ and $\LMM$ are $L\times L$ real, symmetric and non-negative matrices. We note the last two terms of Hamiltonian in Eq.(\ref{eq:Ham}) is new compared to \cite{GuoDario2017} while the dissipator $\Dop$ in Eq.(\ref{eq:dissipator}) remains the same.

\section{solving the master equation}\label{sec:solvemaster}
\subsection{Mapping the density operator into new representations}
As in \cite{GuoDario2017}, first we perform a one-to-one mapping from the density operator basis elements $\vert n_1, n_2, \dots n_L \rangle \langle  n'_1, n'_2, \dots n'_L \vert $ to a state vector basis (with $2L$ sites) which we denote as $\vert n_1, \dots n_L, n'_1, \dots n'_{L} \rangle_{\PA}$. As a result, the operator $\alop_i$ acting on site $i$ to the left of the density matrix is mapped to $\aop_i$ acting on the state vector on the $i$-th site too, while the operator $\alop_i$ acting on the right of the density matrix is mapped to $\adop_{L+i}$ acting on the state vector. We refer to the representation defined by the $2L$ modes $\aop_i$ as the $\PA$ representation.

To enforce the fermionic anti-commutation relations over all the sites, we perform a second mapping from $2L$ modes $\aop_i$ to another set of $2L$ modes $\bop_i$, which we refer to as the $\PB$ representation:
\begin{subequations}
\begin{align}
& \bop_i  = \aop_i, \;\;\;\; \bdop_i = \adop_i \\
&\bop_{L+i} = \Pop\aop_{L+i}, \;\;\;\; \bdop_{L+i} = \adop_{L+i}\Pop,   
\end{align}
\end{subequations}
where $\Pop$ is the parity operator \cite{Prosen2008, GuoDario2017} defined as
\begin{align}
\Pop = e^{\im \pi \sum_{j=1}^{2L}\bdop_j\bop_j}.
\end{align}
The Hamiltonian term in $\PB$ representation can be written as
\begin{align}\label{eq:Hnew}
[ \Hop, \rhoop ]_{\PB} =& \sum_{i,j=1}^L \left( \hb_{i,j}\bdop_i\bop_j + \frac{1}{2} \gb_{i,j}\bdop_i\bdop_j - \frac{1}{2} \gb_{i,j}^{\ast}\bop_i\bop_j \right. \nonumber \\ &- \left.  \hb_{j,i}\bdop_{L+i}\bop_{L+j} - \frac{1}{2} \gb_{i,j}\bop_{L+i}\bop_{L+j} \right. \nonumber \\
 &+ \left. \frac{1}{2} \gb_{i,j}^{\ast}\bdop_{L+i}\bdop_{L+j} \right),
\end{align}
and the dissipative part of Eq.(\ref{eq:Lindblad}) can be written in the $\PB$ representation as
\begin{align}\label{eq:Dnew}
\Dop^{\PB} \vert \rho \rangle_{\PB} =  \sum_{i,j=1}^L & \left( \LMP_{ij} \bdop_i \bdop_{L+j} - \LMP_{ji} \bop_i \bdop_j  + \LMM_{ji} \bop_{L+i} \bop_{j} -\right. \nonumber \\
&\left. \LMM_{ji} \bdop_i \bop_j  - {\LMP_{ij}}  \bdop_{L+i} \bdop_j - {\LMP_{ji}}  \bop_{L+i} \bdop_{L+j}\right. \nonumber \\ - 
&\left.{\LMM_{ji}} \bop_{i}\bop_{L+j}  - 
{\LMM_{ji}} \bdop_{L+i}\bop_{L+j} \right)\vert \rho  \rangle_{\PB} ,
\end{align}
where $\Dop^{\PB}$ is the dissipator $\Dop$ in the ${\PB}$ representation while $\vert \rho \rangle_{\PB}$ the density operator $\rhoop$ in $\PB$ representation. We note that the Lindbladian in the $\PB$ representation, $\Lop^{\PB}$, satisfies 
\begin{align}
\left[\Lop^{\PB}, \Pop\right] = 0
\end{align}
since $\Lop^{\PB}$ is quadratic in operators $\bop$ and $\bdop$, which anti-commute with $\Pop$ as in \cite{GuoDario2017}. As a result, the even parity sector and odd parity sector are separated, and we have dropped the $\Pop$ operator in Eq.(\ref{eq:Dnew}) by assuming that we are working in the even parity sector, namely $\Pop=1$.

\subsection{Master equation in the new representations}
Combining Eqs.(\ref{eq:Hnew}, \ref{eq:Dnew}), and using the more compact notation $\textbf{b}_{1\rightarrow L}$ for the column vector, $\left( \bop_1; \dots \bop_L \right)$, it is possible to rewrite $\Lop^{\PB}$ as
 \begin{align}
\mathcal{L}^{\PB} =& \left(
                                                         \begin{array}{c}
                                                          \textbf{b}_{1\rightarrow L}^{\dagger} \\
                                                          \textbf{b}_{1\rightarrow L} \\
                                                          \textbf{b}_{(L+1)\rightarrow 2L} \\ 
                                                          \textbf{b}_{(L+1)\rightarrow 2L}^{\dagger} \\
                                                         \end{array}
                                                       \right)^t \Gm
             \left(
                                                         \begin{array}{c}
                                                          \textbf{b}_{1\rightarrow L} \\
                                                          \textbf{b}_{1\rightarrow L}^{\dagger} \\
                                                          \textbf{b}_{(L+1)\rightarrow 2L}^{\dagger} \\
                                                          \textbf{b}_{(L+1)\rightarrow 2L} \\
                                                         \end{array}
                                                       \right) \nonumber \\ 
                                            &-\tr(\LMM+\LMP),
\end{align}
where the coefficient matrix $\Gm$ is a $4L\times 4L$ matrix
\begin{align}
\Gm = \left(
                                                       \begin{array}{cccc}
                                                          \bar{\hb} &-\im\gb/2  & \LMP& \textbf{0} \\
                                                          \im\gb^{\ast}/2 &-\bar{\hb}^t  & \textbf{0}& -\LMM \\
                                                          {\LMM}^t & \textbf{0}&  -\bar{\hb}^{\dagger}& \im\gb/2 \\
                                                          \textbf{0} & -{\LMP}^t & -\im\gb^{\ast}/2& \bar{\hb}^{\ast} \\
                                                         \end{array}
                                                       \right).  \label{eq:Gmatrix}  
\end{align}
Here $\bar{\hb} = \frac{1}{2}\left( -\im \hb - {\LMM}^t + \LMP \right)$. Denoting
\begin{align}
\Mm = \left( \begin{array}{cc}
\bar{\hb}  & -\im\gb/2 \\
\im\gb^{\ast}/2 & -\bar{\hb}^t
\end{array} \right)
\end{align}
and
\begin{align}
\Jm = \left( \begin{array}{cc}
\LMP  & \textbf{0} \\
\textbf{0} & -\LMM
\end{array} \right),
\end{align}
we can rewrite $\Gm$ in a more compact form
\begin{align}\label{eq:CompactG}
\Gm = \left( \begin{array}{cc}
\Mm  & \Jm \\
-\Ym \Jm^t \Ym & \Ym \Mm^{\ast} \Ym
\end{array} \right).
\end{align}
Here we have used
\begin{align}
\Ym = -\im \left( \begin{array}{cc}
\textbf{0}  & \textbf{1}_L \\
-\textbf{1}_L & \textbf{0}
\end{array} \right),
\end{align}
where $\textbf{1}_l$ denotes an identity matrix of size $l$. In the following we will also use
matrices
\begin{align}
\Xm = \left( \begin{array}{cc}
\textbf{0}  & \textbf{1}_L \\
\textbf{1}_L & \textbf{0}
\end{array} \right)
\end{align}
and
\begin{align}
\Zm = \left( \begin{array}{cc}
\textbf{1}_L  & \textbf{0} \\
\textbf{0} & -\textbf{1}_L
\end{array} \right).
\end{align}

Now we assume that there exists a transformation 
\begin{align} \label{eq:trans1}
\left( \begin{array}{cccc}
            \textbf{b}_{1\rightarrow L} \\
            \textbf{b}_{1\rightarrow L}^{\dagger}  \\
            \textbf{b}_{(L+1)\rightarrow 2L}^{\dagger}  \\
            \textbf{b}_{(L+1)\rightarrow 2L}  \\
      \end{array} \right) = \Wm \left( \begin{array}{cccc}
            \textbf{c}_{1\rightarrow L}\\
            \textbf{c}_{(L+1)\rightarrow 2L}  \\
            \textbf{c}_{(L+1)\rightarrow 2L}^{\prime}  \\
            \textbf{c}_{1\rightarrow L}^{\prime}   \\
      \end{array} \right),
\end{align}
which perserves the fermionic commutation relation 
\begin{align}\label{eq:commutation}
\left\lbrace \left( \begin{array}{cccc}
            \textbf{c}_{1\rightarrow L}\\
            \textbf{c}_{(L+1)\rightarrow 2L}  \\
            \textbf{c}_{(L+1)\rightarrow 2L}^{\prime}  \\
            \textbf{c}_{1\rightarrow L}^{\prime}   \\
      \end{array} \right), \left( \begin{array}{cccc}
            \textbf{c}_{1\rightarrow L}^{\prime}\\
            \textbf{c}_{(L+1)\rightarrow 2L}^{\prime}  \\
            \textbf{c}_{(L+1)\rightarrow 2L}  \\
            \textbf{c}_{1\rightarrow L}   \\
      \end{array} \right)^t \right\rbrace = \textbf{1}_{4L}.
\end{align}
From Eq.(\ref{eq:trans1}) we have
\begin{align}\label{eq:trans2}
&\left( \begin{array}{cccc}
            \textbf{c}_{1\rightarrow L}^{\prime}\\
            \textbf{c}_{(L+1)\rightarrow 2L}^{\prime}  \\
            \textbf{c}_{(L+1)\rightarrow 2L}  \\
            \textbf{c}_{1\rightarrow L}   \\
      \end{array} \right) \nonumber \\
      &= \left( \begin{array}{cc}
\textbf{0} & \Xm \\
\Xm & \textbf{0}
\end{array} \right)\Wm^{-1} \left( \begin{array}{cc}
\Xm & \textbf{0} \\
\textbf{0} & \Xm
\end{array} \right)\left( \begin{array}{cccc}
            \textbf{b}_{1\rightarrow L}^{\dagger} \\
            \textbf{b}_{1\rightarrow L}  \\
            \textbf{b}_{(L+1)\rightarrow 2L}  \\
            \textbf{b}_{(L+1)\rightarrow 2L}^{\dagger}  \\
      \end{array} \right).
\end{align}
Substituting Eqs.(\ref{eq:trans1}, \ref{eq:trans2}) into Eq.(\ref{eq:commutation}), we get
\begin{align} \label{eq:Winverse}
\Wm^{-1} = \left( \begin{array}{cc}
\textbf{0} & \Xm \\
\Xm & \textbf{0}
\end{array} \right) \Wm^{t} \left( \begin{array}{cc}
\Xm & \textbf{0} \\
\textbf{0} & \Xm
\end{array} \right)
\end{align}
In the following we refer to the new representation defined by $\cop$ as the $\PC$ representation,
using the transformation in Eq.(\ref{eq:trans1}), we get
 \begin{align} \label{eq:LC}
\mathcal{L}^{\PC} = & \left(
                                                         \begin{array}{c}
                                                          \textbf{c}_{1\rightarrow L}^{\prime} \\ 
                                                          \textbf{c}_{L+1\rightarrow 2L}^{\prime} \\
                                                          \textbf{c}_{L+1\rightarrow 2L} \\ 
                                                          \textbf{c}_{1\rightarrow L} \\
                                                         \end{array}
                                                       \right)^t \Wm^{-1}\Gm\Wm
             \left(
                                                         \begin{array}{c}
                                                          \textbf{c}_{1\rightarrow L} \\
                                                          \textbf{c}_{L+1\rightarrow 2L} \\
                                                          \textbf{c}_{L+1\rightarrow 2L}^{\prime} \\
                                                          \textbf{c}_{1\rightarrow L}^{\prime} \\
                                                         \end{array}
                                                       \right) \nonumber \\
                                             &-\tr(\LMM+\LMP),
\end{align}
where $\mathcal{L}^{\PC}$ denotes the Lindbladian $\mathcal{L}$ in the $\PC$ representation. This implies that the problem of finding the normal master modes of the system reduces to diagonalizing the $4L\times 4L$ matrix $\Gm$. 

\subsection{Normal master modes}
The matrix $\Gm$ satisfies two symmetries which imply that, for each eigenvalue $\omega$, there exist also the eigenvalues $-\omega$ and $\pm \omega^*$. These symmetries will be the first step to simplify the problem from solving a $4L\times 4L$ matrix to a $2L\times 2L$  matrix. More in detail, $\Gm$ satisfies
\begin{align}
&\left( \begin{array}{cc}
\Xm & \textbf{0} \\
\textbf{0} & \Xm
\end{array} \right) \Gm \left( \begin{array}{cc}
\Xm & \textbf{0} \\
\textbf{0} & \Xm
\end{array} \right) = -\Gm^t \label{eq:xsymmetry} \\
&\left( \begin{array}{cc}
\textbf{0} & \Ym \\
-\Ym & \textbf{0}
\end{array} \right) \Gm \left( \begin{array}{cc}
\textbf{0} & \Ym \\
-\Ym & \textbf{0}
\end{array} \right) = -\Gm^{\ast} \label{eq:ysymmetry}      
\end{align}
Using Eq.(\ref{eq:xsymmetry}) we find that if 
\begin{align}
\vec{\mathbf{x}} = \left(
\begin{array}{c}
\vec{\mathbf{u}} \\
\vec{\mathbf{v}} \\
\end{array}
\right)
\end{align}
is a right eigenvector of $\Gm$ for the eigenvalue $\omega$, then 
\begin{align}
\vec{\mathbf{x}}^t \left( \begin{array}{cc}
\Xm & \textbf{0} \\
\textbf{0} & \Xm
\end{array} \right) = \left(
\begin{array}{c}
\Xm \vec{\mathbf{u}} \\
\Xm \vec{\mathbf{v}} \\
\end{array}
\right)^t
\end{align}
is a left eigenvector of $\Gm$ to $-\omega$. Moreover, from Eq.(\ref{eq:ysymmetry}) we obtain that  
\begin{align}
\left( \begin{array}{cc}
\textbf{0} & \Ym \\
-\Ym & \textbf{0}
\end{array} \right) \vec{\mathbf{x}}^{\ast} = \left(
\begin{array}{c}
\Ym \vec{\mathbf{v}}^{\ast} \\
-\Ym \vec{\mathbf{u}}^{\ast} \\
\end{array}
\right)
\end{align}
is another right eigenvector of $\Gm$ to $\omega^{\ast}$.
At this point we define a $2L \times 2L$ matrix $\Pm$
\begin{align}\label{eq:defineP}
\Pm = \left(
\begin{array}{cc}
\bar{\Pm} & -\im\gb/2 \\
\im\gb^{\ast}/2 &  \bar{\Pm}^{\ast}
\end{array}
   \right)
\end{align}
with 
\begin{align}
\bar{\Pm} = \bar{\hb} - \LMP = \left( -\im \hb - {\LMM}^t - \LMP \right)/2 
\end{align}
an $L\times L$ matrix. $\Pm$ satisfies the symmetry 
\begin{align}
\Xm \Pm \Xm = \Pm^{\ast}.
\end{align}
Therefore if $\vec{\mathbf{y}}$ is a right eigenvector of $\Pm$ to $\omega$, then $\Xm \vec{\mathbf{y}}^{\ast}$ is another right eigenvector of $\Pm$ to $\omega^{\ast}$. Assuming that $\Pm$ has the eigen-decomposition
\begin{align}\label{eq:eigenP}
\Pm \left(\begin{array}{cc}
\Rm & \Tm^{\ast} \\
\Tm & \Rm^{\ast}
 \end{array} \right) =  \left(\begin{array}{cc}
\Rm & \Tm^{\ast} \\
\Tm & \Rm^{\ast}
 \end{array} \right)  \left(\begin{array}{cc}
\lmp & \textbf{0} \\
\textbf{0} & \lmpc
 \end{array} \right),
\end{align}
then, from Eqs.(\ref{eq:CompactG},\ref{eq:defineP},\ref{eq:eigenP}), it is possible to verify that $ \left(\begin{array}{c}
\Rm \\
\Tm  \\
-\Rm \\
\Tm  
 \end{array} \right)$ and $ \left(\begin{array}{cc}
 -\Tm^{\ast} \\
 -\Rm^{\ast} \\
 \Tm^{\ast} \\
 -\Rm^{\ast}
 \end{array} \right)$ constitute $2L$ right eigenvectors of $\Gm$, corresponding to the $2L$ eigenvalues $\lmp$ and $\lmpc$. From the symmetry in Eq.(\ref{eq:xsymmetry}) we know there exists $2L$ additional eigenvalues which are $-\lmp$ and $-\lmpc$, and we denote the corresponding eigenvectors as  $ \left(\begin{array}{c}
\Am \\
-\Bm  \\
\Cm \\
\Dm  
 \end{array} \right)$ and   $ \left(\begin{array}{c}
\Dm^{\ast} \\
-\Cm^{\ast}  \\
\Bm^{\ast} \\
\Am^{\ast}  
 \end{array} \right)$. With these notations, $\Wm$ can be written as
 \begin{align}\label{eq:W4}
 \Wm = \left(
 \begin{array}{cccc}
 \Rm & -\Tm^{\ast} & \Dm^{\ast} & \Am \\
 \Tm & -\Rm^{\ast} & -\Cm^{\ast} & -\Bm \\
 -\Rm & \Tm^{\ast} & \Bm^{\ast} & \Cm \\
  \Tm & -\Rm^{\ast} & \Am^{\ast} & \Dm \\
 \end{array}
 \right),
 \end{align}
 and then $\Gm$ can be diagonalized as follows
 \begin{align}\label{eq:eigen4}
 \Wm^{-1}\Gm\Wm = \left(
 \begin{array}{cccc}
 \lmp & \textbf{0} & \textbf{0}  & \textbf{0}  \\
 \textbf{0}  & \lmpc & \textbf{0}  & \textbf{0}  \\
 \textbf{0}  & \textbf{0}  & -\lmpc & \textbf{0}  \\
  \textbf{0}  & \textbf{0}  & \textbf{0}  & -\lmp \\
 \end{array}
 \right).
 \end{align}
 Substituting Eq.(\ref{eq:eigen4}) into Eq.(\ref{eq:LC}), also noticing from Eq.(\ref{eq:eigenP}) that
\begin{align}
\sum_{i=1}^L\left(\lambda_{P, i}+\lambda_{P, i}^{\ast}\right) = \tr(\Pm) = - \tr(\LMP+\LMM),
\end{align}
we get the compact expression  
 \begin{align}\label{eq:LCdiag}
 \mathcal{L}^{\PC} =& 2\sum_{i=1}^L \lambda_{P, i}\cdop_i\cop_i + 2\sum_{i=1}^L \lambda_{P, i}^{\ast}\cdop_{L+i}\cop_{L+i}.
 \end{align}
This means that, in order to compute all the $4L$ rapidities, it is sufficient to diagonalize the $2L\times 2L$ matrix $\Pm$ in Eq.(\ref{eq:defineP}).

\section{Computing quadratic observables} \label{sec:observable}
We now show that in order to compute any two-particles observable it is sufficient to solve the Lyapunov equation (\ref{eq:pxxp}). In order to do so we first need to derive the Lyapunov equation, and then we need to show its connection to the two-particles observables. To start, we define
\begin{align}
\Em =& \left( \begin{array}{cc}
\Rm & -\Tm^{\ast} \\
\Tm & -\Rm^{\ast}
\end{array} \right) \\
\Fm =& \left( \begin{array}{cc}
\Dm^{\ast} & \Am \\
-\Cm^{\ast} & -\Bm
\end{array} \right) \\
\lmt =& \left( \begin{array}{cc}
\lmp & \textbf{0} \\
\textbf{0} & \lmpc
\end{array} \right).
\end{align}
We can see that $\Xm \Em \Xm = -\Em^{\ast}$. With these definitions, we can write $\Wm$ as
\begin{align}
\Wm = \left( \begin{array}{cc}
\Em & \Fm \\
-\Zm \Em & -\im\Ym \Fm^{\ast} \Xm
\end{array} \right).
\end{align}
Using Eq.(\ref{eq:Winverse}), we get
\begin{align}
\Wm^{-1} = \left( \begin{array}{cc}
\Xm \Fm^t \Xm & \Fm^{\dagger}\Zm \\
-\Em^{\dagger} & -\Em^{\dagger}\Zm
\end{array} \right).
\end{align}
Then from $\Wm^{-1}\Wm = \textbf{1}_{4L}$, we get
\begin{align}
\left( \begin{array}{cc}
\Xm \Fm^t \Xm \Em - \Fm^{\dagger}\Em & \Xm \Fm^t \Xm \Fm - \Fm^{\dagger}\Xm \Fm^{\ast}\Xm \\
\textbf{0} & -\Em^{\dagger}\Fm + \Em^{\dagger}\Xm \Fm^{\ast}\Xm 
\end{array} \right) = \textbf{1}_{4L},
\end{align}
from which we get two independent matrix equations
\begin{align}
&\Xm \Fm^t \Xm \Em - \Fm^{\dagger}\Em = \textbf{1}_{2L} \rightarrow \Fm^{\dagger} = \Xm \Fm^t \Xm - \Em^{-1} \label{eq:indep1} \\
&\Xm \Fm^t \Xm \Fm - \Fm^{\dagger}\Xm \Fm^{\ast}\Xm = \textbf{0}
\end{align}
Now we rewrite Eq.(\ref{eq:eigen4}) in terms of $\Em, \Fm, \lmt$
\begin{align}
&\left( \begin{array}{cc}
\Mm  & \Jm \\
-\Ym \Jm^t \Ym & \Ym \Mm^{\ast} \Ym
\end{array} \right)\left( \begin{array}{cc}
\Em & \Fm \\
-\Zm \Em & -\im\Ym \Fm^{\ast} \Xm
\end{array} \right) \nonumber \\ 
=&\left( \begin{array}{cc}
\Em & \Fm \\
-\Zm \Em & -\im\Ym \Fm^{\ast} \Xm
\end{array} \right)\left(\begin{array}{cc}
\lmt & \textbf{0}  \\
\textbf{0} & -\lmtc \\
 \end{array} \right),
\end{align}
from which we have
\begin{align}
& \Mm\Em - \Jm\Zm \Em = \Em \lmt \\
& \Mm\Fm - \im \Jm \Ym \Fm^{\ast}\Xm = -\Fm \lmtc
\end{align}
Since $\Fm^{\ast} = \Xm \Fm\Xm - {\Em^t}^{-1}$, we have
\begin{align}
\Mm\Fm - \im\Jm\Ym(\Xm \Fm\Xm-{\Em^t}^{-1})\Xm = -\Fm\lmtc.
\end{align}
from which we get
\begin{align}
(\Mm - \Jm\Zm)\Fm + \im\Jm\Ym {\Em^t}^{-1}\Xm = -\Fm \Xm \lmt \Xm
\end{align}
Since $\Mm -\Jm\Zm = \Pm$, we have
\begin{align}
\Pm\Fm\Xm \Em^t + \im\Jm \Ym = -\Fm\Xm \lmt\Em^t = -\Fm\Xm \Em^t \Pm^t,
\end{align}
therefore we get
\begin{align}
-\Pm \Fm \Xm \Em^t \Xm - \Fm\Xm \Em^t \Xm \Pm^{\dagger} = \Jm \Zm.
\end{align}
Denoting $\Omegam = -\Fm\Xm \Em^t \Xm$, we get the equation for $\Omegam$
\begin{align}\label{eq:pxxp}
\Pm\Omegam+\Omegam\Pm^{\dagger} = \Jm\Zm.
\end{align}
This is Lyapunov equation, which can be solved with methods which scale as $O(L^3)$ \cite{Lyap1, Lyap2}. We should also stress that for quadratic open systems, it is often possible reduce the analysis of the problem in equations of this form, see for example \cite{ZunkovicProsen2010, BanchiZanardi2014}.  

Now we show that Eq.(\ref{eq:pxxp}) is relevant to compute observables. We start by demonstrating that
\begin{align}\label{eq:annihilation}
{}_{\PA}\langle \id \vert \cdop_k = 0, \quad \forall 1 \leq k \leq 2L,
\end{align}
where ${}_{\PA}\langle \id \vert$ is the transpose of the identity operator in the $\PA$ representation, $\vert \id \rangle = \sum_{i_1, i_2, \dots, i_L}\vert i_1, i_2, \dots, i_L, i_1, i_2, \dots, i_L \rangle_{\PA}$. From the inverse of Eq.(\ref{eq:trans1}) we have
\begin{subequations}
\begin{align}
\cop_{i}^{\prime} &= \sum_{k=1}^L\left( \Tm^t_{ik}\bop_k + \Rm^t_{ik}\bdop_k + \Tm^t_{ik}\bdop_{L+k} - \Rm^t_{ik}\bop_{L+k} \right) \nonumber \\
& =\sum_{k=1}^L\left[ \Tm^t_{ik}\left(\bop_k +\bdop_{L+k} \right) + \Rm^t_{ik}\left(\bdop_k-\bop_{L+k}\right) \right]
 \\
\cop_{L+i}^{\prime} &= \sum_{k=1}^L \left(-\Rm^{\dagger}_{ik}\bop_k - \Tm^{\dagger}_{ik}\bdop_k - \Rm^{\dagger}_{ik}\bdop_{L+k} + \Tm^{\dagger}_{ik}\bop_{L+k} \right) \nonumber \\
&=\sum_{k=1}^L \left[-\Rm^{\dagger}_{ik}\left(\bop_k+\bdop_{L+k}\right) - \Tm^{\dagger}_{ik}\left(\bdop_k-\bop_{L+k}\right) \right].
\end{align}
\end{subequations}
Therefore to prove Eq.(\ref{eq:annihilation}), it is sufficient to prove that for any $1\leq k \leq L$,
\begin{align}\label{eq:identity1}
{}_{\PA}\langle \id \vert \left(\bop_k + \bdop_{L+k}\right) = 0 
\end{align}
and
\begin{align}\label{eq:identity2}
{}_{\PA}\langle \id \vert \left(\bdop_k - \bop_{L+k}\right) = 0,
\end{align}
which has already been proved in \cite{GuoDario2017}. Now using Eq.(\ref{eq:trans1}) we get
\begin{subequations} \label{eq:btoc}
\begin{align} 
\bop_i &= \sum_{k=1}^L\left( \Rm_{ik}\cop_k - \Tm_{ik}^{\ast}\cop_{L+k} + \Dm_{ik}^{\ast}\cop_{L+k}^{\prime} + \Am_{ik}\cop_k^{\prime} \right) \\
\bdop_i &= \sum_{k=1}^L\left( \Tm_{ik}\cop_k - \Rm_{ik}^{\ast}\cop_{L+k} - \Cm_{ik}^{\ast}\cop_{L+k}^{\prime} - \Bm_{ik}\cop_k^{\prime} \right).
\end{align}
\end{subequations}
We can thus write 
\begin{align}
\Om =& \tr\left(\left(
\begin{array}{c}
\textbf{b}^{\dagger} \\
\textbf{b} \\
\end{array}
\right) \left(
\begin{array}{c}
\textbf{b} \\
\textbf{b}^{\dagger} \\
\end{array}
\right)^t \rhoop\right) \nonumber \\
=& {}_{\PA}\langle \id \vert \left(
\begin{array}{c}
\textbf{b}^{\dagger} \\
\textbf{b} \\
\end{array}
\right) \left(
\begin{array}{c}
\textbf{b} \\
\textbf{b}^{\dagger} \\
\end{array}
\right)^t \vert \rho_{\rm ss} \rangle.
\end{align}
Substituting Eqs.(\ref{eq:btoc}) into the above equation, and using Eq.(\ref{eq:annihilation}), we get
\begin{align}
\Om =& \left(\begin{array}{cc}
\Tm\Am^t - \Rm^{\ast}\Dm^{\dagger} & -\Tm\Bm^t + \Rm^{\ast}\Cm^{\dagger} \\
\Rm\Am^t - \Tm^{\ast}\Dm^{\dagger} & -\Rm\Bm^t + \Tm^{\ast}\Cm^{\dagger}
\end{array}
\right) \nonumber \\
=&  \Xm \Em \Xm \Fm^t = -\Omegam^t .
\end{align}
Therefore, the quadratic observables $\Om$ can be determined by solving Eq.(\ref{eq:pxxp}).

\section{Solutions for a boundary driven XY model}\label{sec:solveXY}
In the following we apply our method to a boundary driven $XY$ model. The Lindblad equation in this case is
\begin{align}
\Lop_{\rm XY} (\rhoop) = -\im [\Hop_{\rm XY}, \rhoop] + \Dop_{\rm XY}(\rhoop),
\end{align}
with
\begin{align}
\Hop_{\rm XY} =& \frac{J\left(1+\gamma\right)}{2}\sum_{i=1}^{L-1}\sgx_i\sgx_{i+1} \nonumber \\
&+ \frac{J\left(1-\gamma\right)}{2}\sum_{i=1}^{L-1}\sgy_i\sgy_{i+1} + h_z \sum_{i=1}^L\sgz_i,
\end{align}
and
\begin{align}
\Dop_{\rm XY}(\rhoop) = \sum_{l=1,L}&\left[\Lambda^+_{l}(2\sgp_{l}\rhoop \sgm_{l}-\{\sgm_{l}\sgp_{l}, \rhoop\}) \right. \\ &+ \left.  \Lambda^-_{l}(2\sgm_{l}\rhoop \sgp_{l}- \{\sgp_{l}\sgm_{l}, \rhoop \}) \right], \label{eq:DissLC}    
\end{align}
Applying Jordan-Wigner transformation \cite{JordanWigner1928, LiebMattis1961}, the $XY$ chain can be mapped into a fermionic chain
\begin{align}
\Hop_F =& J\sum_{i=1}^{L}\left[\alop_i\aldop_{i+1}+\alop_{i+1}\aldop_i + \gamma\left(\alop_i\alop_{i+1}+\aldop_{i+1}\aldop_i\right)\right] \nonumber \\
&+h_z\sum_{i=1}^L(2\aldop_i\alop_i-1),
\end{align}
The dissipator can also be mapped into fermionic representation $\PB$ as in \cite{GuoDario2017}, after which we can read the $L\times L$ matrices $\bar{\Pm}$ and $\gb$ with non-zero elements
\begin{align}
&\Pm_{l, l+1} = \Pm_{l+1, l} = -\frac{\im J}{2}; \quad \forall 1 \leq l < L \\
&\Pm_{l, l} = -\im h_z; \quad \forall 1 < l < L  \\
&\Pm_{1, 1} = -\im h_z - \frac{\Gamma_1}{2}, \Pm_{L, L} = -\im h_z - \frac{\Gamma_L}{2}, 
\end{align}
where $\Gamma_1 = \Lambda_1^+ + \Lambda_1^-$ and $\Gamma_L = \Lambda_L^+ + \Lambda_L^-$,
and 
\begin{align}
\gb_{l, l+1} = J\gamma, \quad \gb_{l+1, l} = -J\gamma; \quad \forall 1 \leq l < L.
\end{align}
$\bar{\Pm}$ is a tridiagonal bordered Toeplitz matrix, and $\gb$ is an anti-symmetric tridiagonal Toeplitz matrix. In the isotropic case of $\gamma=0$, $\Pm$ is block diagonal with $\bar{\Pm}$ and $\bar{\Pm}^{\ast}$ on its diagonal, $\bar{\Pm}$ has shown to be analytically diagonalizable in \cite{Yueh2005, Kouachi2006, Willms2008}, something which we exploited in \cite{GuoDario2017}.\\ 

Here we show analytical solutions for arbitrary $J, \gamma$, provided $h_z=0$. Eq.(\ref{eq:eigenP}) for $\Pm$ can be expanded into two independent equations
\begin{subequations}\label{eq:eigenPsub}      
\begin{align}
&\bar{\Pm}\Rm - \frac{\im \gb\Tm}{2} = \Rm\lmp \\ 
&\bar{\Pm}^{\ast}\Tm + \frac{\im \gb\Rm}{2} = \Tm\lmp.
\end{align}
\end{subequations}
We then introduce two $L\times L$ diagonal matrices $\Kpm$ with the diagonal elements
\begin{align}
\Kp_{i, i} &= (-1)^{i+1} \\
\Km_{i, i} &= (-1)^i,
\end{align}
for $1 \leq i \leq L$. For the $XY$ chain the following relations
\begin{align}
&\Kpm \Kpm = \id_L \\
&\Kpm \bar{\Pm} \Kpm = \bar{\Pm}^{\ast} \\
&\Kpm \gb \Kpm = -\gb 
\end{align}
are valid. Using them we can solve Eqs.(\ref{eq:eigenPsub}) with the ansatz
\begin{align}\label{eq:ansatz}
\Tm = \Kpm \Rm.
\end{align}
Substituting Eq.(\ref{eq:ansatz}) into Eqs.(\ref{eq:eigenPsub}), we get a single equation 
\begin{align}
\left(\bar{\Pm}- \frac{\im \gb \Kpm }{2}\right) \Rm = \Rm\lmp.
\end{align}
Therefore, to diagonalize the boundary driven $XY$ chain with $0$ magnetic field, one only needs to diagonalize two $L\times L$ matrices 
\begin{align}
\Qm^{\pm} = \bar{\Pm}- \im \gb \Kpm/2. \label{eq:Qpm}   
\end{align}
Moreover, $\Qm^{\pm}$ is a tridiagonal bordered 2-Toeplitz matrix, whose characteristic determinant is
\begin{align}\label{eq:exactodd}
\Delta_L = &\frac{(d_1d_2)^{m-1}}{\sin(\theta)}\left[d_1d_2\left(\frac{\Gamma_1+\Gamma_L}{2}-\lambda\right)\sin(m+1)\theta \right. \nonumber \\ &- \left. \left(\frac{\Gamma_1\Gamma_L\lambda}{4} + \frac{d_1^2\Gamma_1 + d_2^2\Gamma_L}{2}\right)\sin(m\theta)\right],
\end{align}
when $L=2m+1$ is odd and
\begin{align}\label{eq:exacteven}
\Delta_L = &\frac{(d_1d_2)^{m-1}}{\sin(\theta)}\left[\left(\frac{\Gamma_1\Gamma_L}{4}+d_2^2+\frac{(\Gamma_1+\Gamma_L)\lambda}{2}\right)\sin(m\theta) \right. \nonumber \\ 
&+ \left. d_1d_2\sin(m+1)\theta + \frac{\Gamma_1\Gamma_L}{4} \frac{d_1}{d_2} \sin(m-1)\theta \right]
\end{align}
when $L=2m$ is even (see Eqs.(4.a, 4.b) in \cite{Kouachi2006}). Here the eigenvalues $\lambda$ and $\theta$ are related by
\begin{align}
\lambda^2 = d_1^2+d_2^2 + 2d_1d_2\cos(\theta).   \label{eq:lambda}  
\end{align}
And $d_1, d_2$ are define as
\begin{align}
&d_1 = -\im J\frac{1 \mp\gamma}{2} \\
&d_2 = -\im J\frac{1 \pm\gamma}{2} 
\end{align}
for $\Qm^{\pm}$, respectively. Therefore, the diagonalization of matrices $\Qm^{\pm}$ are reduced to solving the scalar trigonometric equations Eqs.(\ref{eq:exactodd}, \ref{eq:exacteven}) in the complex number $\theta$. For an Ising chain, which corresponds to $\gamma=1$, we have $d_1d_2=0$. In this special case, Eqs.(\ref{eq:exactodd}, \ref{eq:exacteven}) have closed analytic solutions and all the allowed eigenvalues are (see Proposition 4.2 in \cite{Kouachi2006})
\begin{align}
\left\lbrace \pm\im J, -\frac{\Gamma_{1,L}}{2},  -\frac{\Gamma_{1,L}}{4}\pm\frac{\sqrt{\Gamma_{1,L}^2-16J^2}}{4} \right\rbrace . \label{eq:solutions}
\end{align}
It is interesting to point out that none of these eigenvalues depend on $L$, which means the Lindbladian has a constant relaxation gap irrespective of the system size $L$.\\  

\begin{figure}
\includegraphics[width=\columnwidth]{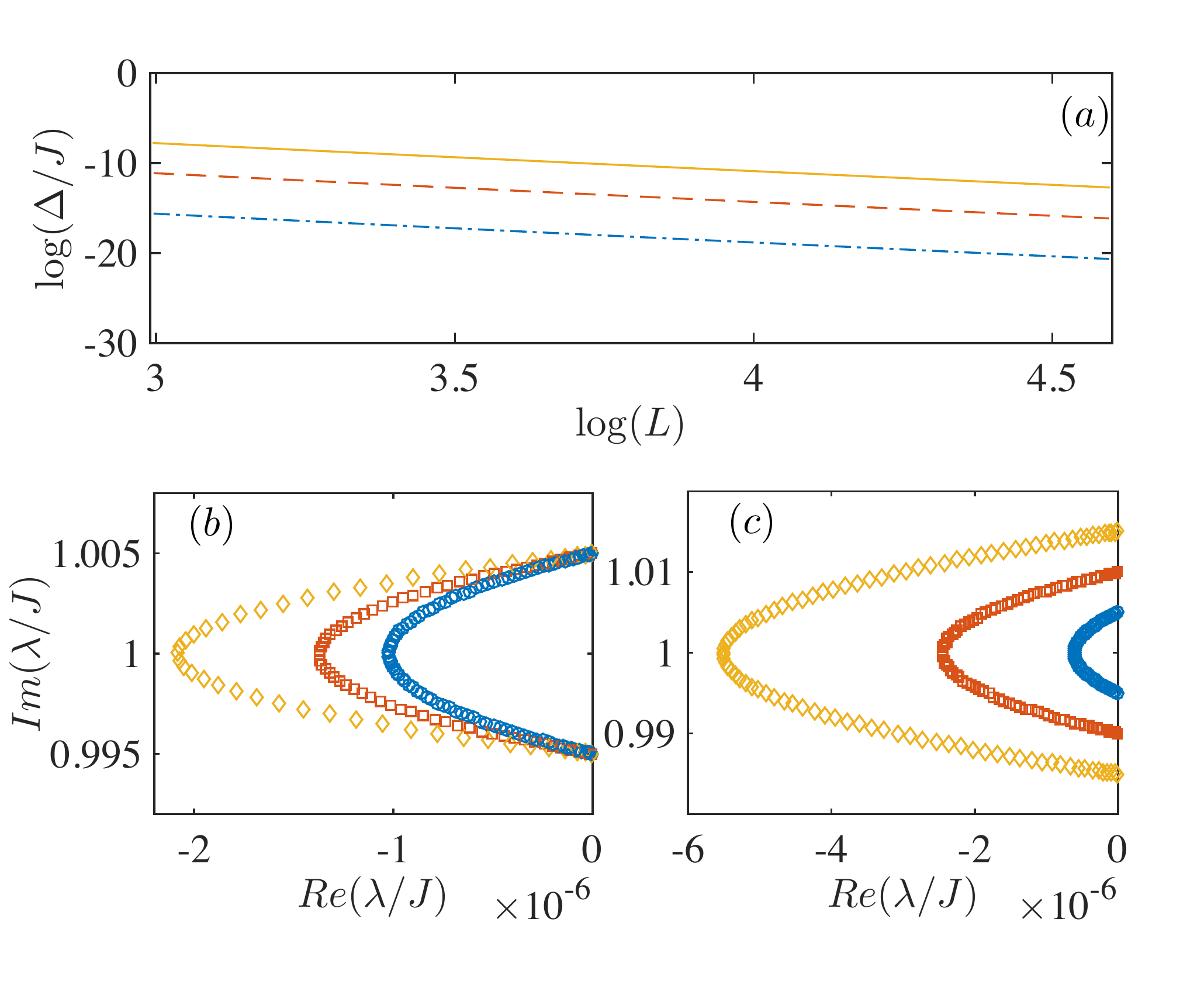}
\caption{ (a) Relaxation gap $\Delta$ versus system size $L$ for different values of magnetic field $h_z$. (b) Real and imaginary part of the rapidities $\lambda$ closest to the real axis and with positive imaginary part, for $h_z/J=0.01$ and different system sizes: yellow diamonds for $L=50$, red squares for $L=75$ and blue circles for $L=100$. (c) Real and imaginary part of the rapidities $\lambda$ closest to the real axis and with positive imaginary part, for $L=100$ and different magnetic fields: yellow diamonds for $h_z/J=0.03$, red squares for $h_z/J=0.02$ and blue circles for $h_z/J=0.01$. Common parameters to the three panels are $\gamma=1$, $\Gamma_1=1$, $\Gamma_L=1$.   	
} \label{fig:fig1} 
\end{figure}

We should now compare this result with numerical solutions and with the perturbative expression in \cite{Prosen2008}. In this study it was found that for $h_z \ne 0$ the relaxation gap scales as $1/L^3$ (see Eq.(85) of \cite{Prosen2008}). There is no contrast between this result and ours, because for $h_z=0$ the predicted relaxation gap (real part of the slowest decaying mode) also vanishes. In Fig.\ref{fig:fig1}(a) we show how the relaxation gap $\Delta$ scales with the system size for different values of the magnetic field $h_z$, computed by diagonalizing numerically the $2L\times 2L$ matrix $\Pm$ in Eq.(\ref{eq:defineP}). The scaling follows a power-law well approximated by $L^{-3}$, and the gap decreases when $h_z$ is smaller. We also investigate the real and imaginary parts of the rapidities $\lambda$. For small $h_z$, the rapidities with the smallest real part are found near $\pm\im J$ and approach these values for larger system size $L$ and as $h_z$ tends to $0$. This is shown in Fig.\ref{fig:fig1}(b-c) where we respectively increase the system size $L$ or decrease $h_z$.     

We note here that the presence of slow decaying modes which have a large imaginary part can result, when computing two-time observables on the steady state, in long-lasting periodic motions which break the continuous time-translation symmetry, thus resulting in a time crystal \cite{Wilczek2012, WatanabeOshikawa2015, SachaZakrzewski2018}. For studies using two-time correlations to investigate time crystal in dissipative systems see for example \cite{WangPoletti2018, TuckerRey2018}.

\section{conclusions}\label{sec:summary}
In this work we have studied the steady state of dissipatively boundary driven fermionic quadratic system in which the particle number is not conserved, i.e. an $XY$ chain. We have shown that, not only it is possible to convert the problem of computer all the relaxation rates and normal master mode of the Lindblad master equation to diagonalizing an $L\times L$ matrix (where $L$ is the number of spins), but also that the matrix has a particular structure (it is a tridiagonal bordered $2-$Toeplitz matrix) which can then be solved as a scalar trigonometric equation. 
Moreover, for the special case of the Ising chain we find explicit analytical solutions which are independent of the system size $L$.   
The method here presented can be useful to study both the time evolution (since it gives access to all the normal master modes and rapidities) and steady states for open quadratic fermionic systems far from equilibrium even when the total number of particles is not conserved. Note that, once the problem is brought into a $L\times L$ matrix of bordered $2-$Toeplitz form, it may also be possible to find other further analytical solutions to Eqs.(\ref{eq:exactodd}, \ref{eq:exacteven}), as well as the expressions for the eigenvectors, by referring to, for example, references \cite{Yueh2005, Kouachi2006, Willms2008, Fonseca2007}.   

\begin{acknowledgments} 
D.P. acknowledges discussions with A.M. Rey. D.P. also acknowledges support from the Ministry of Education of Singapore AcRF MOE Tier-II (Project No. MOE2016-T2-1-065). C.G. acknowledges support from National Natural Science Foundation of China (11504430).
\end{acknowledgments}


\begin{thebibliography}{99}
\bibitem{Gardiner2000} C. W. Gardiner, and P. Zoller, \emph{\bibinfo{title}{Quantum Noise}} (2000).

\bibitem{BenentiWhitney2017} G. Benenti, G. Casati, K. Saito, and R.S. Whitney, Physics Reports {\bf 694}, 1 (2017)             

\bibitem{NielsenChuang2000} M. A. Nielsen, and I. L. Chuang, \emph{\bibinfo{title}{Quantum Computation and Quantum Information}}, \bibinfo{publisher}{Cambridge University Press}, \bibinfo{address}{Cambridge} (2000).


\bibitem{GoriniSudarshan1976} V. Gorini, A. Kossakowski, and E.C.G. Sudarshan, J. Math. Phys. {\bf 17}, 821 (1976).     
\bibitem{Lindblad1976} G. Lindblad, Comm. Math. Phys. {\bf 48}, 119 (1976). 

\bibitem{Prosen2011a} T. Prosen, Phys. Rev. Lett {\bf106}, 217206 (2011).
\bibitem{Prosen2011b} T. Prosen, Phys. Rev. Lett {\bf107}, 137201 (2011).
\bibitem{Prosen2014} T. Prosen, Phys. Rev. Lett {\bf112}, 030603 (2014).
\bibitem{Popkov2013b} D. Karevski, V. Popkov, and G.M. Schutz, Phys. Rev. Lett {\bf110}, 047201 (2013). 
\bibitem{Popkov2014} V. Popkov, D. Karevski, and G.M. Schutz, Phys. Rev. E {\bf88}, 062118 (2013). 
\bibitem{Prosen2015} T. Prosen, J. Phys. A: Math. Theor. {\bf 48}, 373001 (2015).


\bibitem{Znidaric2010} M. Znidaric, J. Stat. Mech.: Theory Exp. (2010) L05002.
\bibitem{MedvedyevaProsen2016} M. V. Medvedyeva, F. H. L. Essler, and T. Prosen, Phys. Rev. Lett. {\bf 117}, 137202 (2016).

\bibitem{Manzano2012} D. Manzano, M. Tiersch, A. Asadian, and H. J. Briegel, Phys. Rev. E {\bf 86}, 061118 (2012).

\bibitem{Manzano2013} A. Asadian, D. Manzano, M. Tiersch, and H. J. Briegel, Phys. Rev. E {\bf 87}, 012109 (2013).

\bibitem{Prosen2008} T. Prosen, New J. Phys. {\bf16}, 063062 (2008).

\bibitem{Prosen2010} T. Prosen, J.Stat. Mech: Theory Exp. (2010){\bf P}07020.

\bibitem{ProsenSeligman2012} T. Prosen, and T. H. Seligman, J. Phys. A {\bf 43}, 392004 (2010).     

\bibitem{GuoDario2017} C. Guo, and D. Poletti, Phys. Rev. A {\bf 95}, 052107 (2017).

\bibitem{Yueh2005} W.C. Yueh, Appl.Math. E-Notes {\bf 5}, 66-74 (2005).
\bibitem{Kouachi2006} S. Kouachi, Electron. J. Lin. Alg. {\bf 15}, 115 (2006). 
\bibitem{Willms2008} A.R. Willms, SIAM J. Matrix Anal. Appl. {\bf 30}, 639 (2008). 
\bibitem{Fonseca2007} C. M. da Fonseca, Appl. Math. Sci., {\bf 1}, 59-67 (2007). 

\bibitem{Lyap1} R. H. Bartels, and G. W. Stewart, Commun. ACM {\bf 15}, 820 (1972). 
\bibitem{Lyap2} G. H. Golub, S. Nash, and C. F. Van Loan, IEEE Trans. Autom. Control {\bf 24}, 909 (1979). 

\bibitem{ZunkovicProsen2010} B. \v{Z}unkovi\v{c}, and T. Prosen, Journal of Statistical Mechanics: Theory and Experiment P08016 (2010). 
\bibitem{BanchiZanardi2014} L. Banchi, P. Giorda, and P. Zanardi, Phys. Rev. E {\bf 89}, 022102 (2014).                 

\bibitem{JordanWigner1928} P. Jordan, and E. Wigner, Z. Physik {\bf 47}, 631 (1928).    
\bibitem{LiebMattis1961} E. Lieb, T. Schultz, and D. Mattis, Ann. of Phys. {\bf 16}, 407 (1961). 

\bibitem{Wilczek2012} F. Wilczek, Phys. Rev. Lett. {\bf 109}, 160401 (2012). 
\bibitem{WatanabeOshikawa2015} H. Watanabe, and M. Oshikawa, Phys. Rev. Lett. {\bf 114}, 251603 (2015).       
\bibitem{SachaZakrzewski2018} K. Sacha, and J. Zakrzewski, Rep. Prog. Phys. {\bf 81}, 016401 (2018). 

\bibitem{WangPoletti2018} R.R.W. Wang, B. Xing, G.G. Carlo, and D. Poletti, Phys. Rev. E {\bf 97}, 020202 (2018).     
\bibitem{TuckerRey2018} K. Tucker, B. Zhu, R.J. Lewis-Swan, J. Marino, F. Jimenez, J.G. Restrepo, and A.M. Rey, arxiv:1805.03343.   

   

\end{thebibliography}
\end{document}